\documentclass[12pt]{article}
\usepackage{}
\begin{document}
\date{}
\title{\textbf{Gauge Symmetries on $\theta$-Deformed Spaces}}
\author{{Rabin Banerjee}\thanks{E-mail: rabin@bose.res.in} \ and {Saurav Samanta}\thanks{E-mail: saurav@bose.res.in}\\
\\\textit{S.~N.~Bose National Centre for Basic Sciences,}
\\\textit{JD Block, Sector III, Salt Lake, Kolkata-700098, India}}
\maketitle
                                                                                
\begin{quotation}
\noindent \normalsize 
A Hamiltonian formulation of gauge symmetries on noncommutative ($\theta$ deformed) spaces is discussed. Both cases- star deformed gauge transformation with normal coproduct and undeformed gauge transformation with twisted coproduct- are considered. While the structure of the gauge generator is identical in either case, there is a difference in the computation of the graded Poisson brackets that yield the gauge transformations. Our analysis provides a novel interpretation of the twisted coproduct for gauge transformations.
\end{quotation}
\section{Introduction}
The analysis of gauge symmetries in theories defined on usual commutative space is quite familiar, either in the Lagrangian or Hamiltonian formalisms. In the later formalism, for instance, there is a definite method\cite{henn,rothe,rothe1} of obtaining the gauge generator, based on Dirac's\cite{dirac} conjecture that it has to be a linear combination of the first class constraints. Poisson bracketing the generator with the variables then yields their specific gauge transformations.

In this paper we provide a systematic Hamiltonian analysis of gauge symmetries in noncommutative theories; i. e. theories defined on a noncommutative space where the usual pointwise multiplication is replaced by the star multiplication. Our motivation stems from recent analysis\cite{wess,vas,chai,jwess,gaume,saurav} which show that, in extending gauge symmetries from the usual (commutative) to the noncommutative realm, one is faced with a choice. Either gauge transformations are deformed in such a way that the usual coproduct (Leibniz) rule is preserved or the standard commutative space gauge transformations are retained at the expense of twisting the normal coproduct rule. While the former is referred as star deformed gauge symmetry, the latter is called twisted gauge symmetry. These notions have also been considered in the context of gravity\cite{gaume}. Because of this ambiguity it is clear that extending the concepts of gauge generators and transformations from the commutative to the noncommutative realm is quite nontrivial.

In this paper we analyse both types of gauge symmetry in the Hamiltonian formulation, complementing the Lagrangian approach done by us \cite{saurav}. As a specific model, the noncommutative Yang Mills action coupled to fermionic matter has been taken. The first class constraints of the theory are identified. The gauge generator is constructed by taking an appropriate combination of these constraints. Poisson bracketing the generator with the gauge or matter variables leads to the star deformed gauge transformations. Subsequently by providing a ``twist" to the Poisson brackets, the twisted gauge transformations are obtained. This twist is dictated by a novel interpretation of the twisted coproduct of gauge transformations. We find that the twisted coproduct is the normal coproduct with the stipulation that the gauge parameter is pushed outside the star operation at the end of all computations. 

The paper is organized as follows. In section 2 we discuss a general formulation to obtain the gauge generator and the gauge transformations in a noncommutative space framework. Section 3 is dedicated for the analysis of star deformed gauge transformation. In section 4 we concentrate on the twisted gauge transformations. The issues related to the twisted coproduct rule is discussed in detail in this section. Finally section 5 is for conclusions. 
\section{General Formulation}
The gauge symmetry of a system can be analysed either by the Lagrangian approach\cite{gitman} where the starting point is the gauge identity of that system expressed in terms of the Euler derivatives or by the Hamiltonian approach, based on Dirac's conjecture that the generators of the gauge transformation are given by a linear combination of the first class constraints. Here we concentrate on the second approach for obtaining the gauge transformations of the fields on a noncommutative space.

Let us first briefly mention the formulation for a general field theoretical model defined on a noncommutative space. The results are basically appropriate star deformation of the commutative space results. The star product is defined as usual,
\begin{eqnarray}
(f*g)(x)={\textrm{exp}}\left(\frac{i}{2}\theta^{\mu\nu}\partial_{\mu}^x\partial_{\nu}^y\right)f(x)g(y)|_{x=y}
\label{str}
\end{eqnarray}
where $\theta^{\mu\nu}$ is a constant two index antisymmetric object. We consider a system with a canonical Hamiltonian $H_c$ and a set of first class constraints $\Phi_{a}\approx0${\footnote{The weak equality in $\Phi_{a}\approx0$ implies that all Poisson brackets involving $\Phi_{a}$ have to be calculated first and then only the constraints can be imposed. In contrast, a strong equality $A=0$ implies that $A$ (obviously) has vanishing Poisson bracket with all the phase space variables.}} which satisfy the involutive algebra
\begin{eqnarray}
&&\{H_c,\Phi_a(x)\}=\int \textrm{d}y \ V^b_a(x,y)*\Phi_b(y),
\label{ve}\\
&&\{\Phi_a(x),\Phi_b(y)\}=\int \textrm{d}z \ C^c_{ab}(x,y,z)*\Phi_c(z)
\label{cc}
\end{eqnarray}
where $V$ and $C$ are structure functions which, in general, depend on the field variables. The constraints coming directly from the definition of canonical momentum are named primary constraints and that obtained from their time consistency (the Poisson brackets between the Hamiltonian and the constraints should be weakly zero) are called secondary constraints.

For such a system the total Hamiltonian is given by the sum of the canonical Hamiltonian and a linear combination of the primary first class constraints.
\begin{eqnarray}
H_T = H_c + \int \textrm{d}x \ v^{a_1}(x)*\Phi_{a_1}(x)
\end{eqnarray}
Here $v^{a_1}$ are Lagrange multipliers. The equations of motion in the Hamiltonian formulation are now given by
\begin{eqnarray}
\dot q_i(x) = \{q_i(x),H_T\} = \{q_i(x),H_c\} + \int\textrm{d}y \ v^{a_1}(y)*\{q_i(x),\Phi_{a_1}(y)\}
\label{eof}
\end{eqnarray}
with the constraint equation
\begin{eqnarray}
\Phi_{a_1} \approx 0.
\end{eqnarray}
The generator of the system, according to Dirac's algorithm is a linear combination of all the first class constraints,
\begin{eqnarray}
&&G=\int \textrm{d}x \ \epsilon^a(x)*\Phi_a(x).
\label{G6}
\end{eqnarray}
The point to emphasise is that all components of the gauge parameters $\epsilon^a$ are not independent. The number of independent $\epsilon '$s is given by the number of independent primary first class constraints (labeled by `$a_1$'). To find the conditions among these parameters, we review the method used earlier by one of us \cite{rabin} which is an adaptation of the commutative space approach discussed in\cite{henn,rothe,rothe1}.

An infinitesimal gauge transformation of a variable is given by the Poisson bracket, defined below, between the variable $F$ and the gauge generator $G$,
\begin{eqnarray}
\delta F(x) =\int \textrm{d}y \  \epsilon_a(y)*\{F(x),\Phi^a(y)\}
\label{gt}
\end{eqnarray}
The point is that in demonstrating the invariance of the action under some variation or in the derivation of the Euler-Lagrange equation of motion from the action principle, one requires the commutativity of that ($\delta$) variation with the time differentiation. In the Hamiltonian framework also we impose that requirement,
\begin{eqnarray}
\delta \frac{d}{dt}q_i =  \frac{d}{dt}\delta  q_i
\end{eqnarray}
where the time differentiation is defined in (\ref{eof}) and the $\delta q$ variation in (\ref{gt}). From these equations we obtain
\begin{eqnarray}
\delta \dot q_i(x) &=& \int \textrm{d}z \ \epsilon^a(z)*\{\{q_i(x),H_c\},\Phi_a(z)\} +\nonumber\\
&&\int \int\textrm{d}y \ \textrm{d}z \ \epsilon^b(z)*v^{a_1}(y)*\{\{q_i(x),\Phi_{a_1}(y)\},\Phi_b(z)\}+\nonumber\\
&&\int\textrm{d}y \ \delta v^{a_1}(y)*\{q_i(x),\Phi_{a_1}(y)\}.
\label{deldot}
\end{eqnarray}
Similarly we can write
\begin{eqnarray}
\frac{d}{dt}\delta q_i(x)& =& \int\textrm{d}y \  \epsilon^a(y)*\{\{q_i(x),\Phi_a(y)\},H_c\}\nonumber\\&&
+ \int\textrm{d}y \  \textrm{d}z \ \epsilon^a(y)*v^{a_1}(z)*\{\{q_i(x),\Phi_a(y)\},\Phi_{a_1}(z)\}\nonumber\\&&
+\int\textrm{d}y \  \frac{d\epsilon^a}{dt}(y)*\{q_i(x),\Phi_a(y)\}.
\label{dotdel}
\end{eqnarray}
Equating (\ref{deldot}) and (\ref{dotdel}) and using the Jacobi identity we get
\begin{eqnarray}
&&\int\textrm{d}z \ \epsilon^a(z)*\{\{H_c,\Phi_a(z)\}, q_i\}\nonumber\\ 
&&+\int\textrm{d}y \ \textrm{d}z \  \epsilon^a(z)*
v^{a_1}(y)*\{\{\Phi_{a_1}(y),\Phi_{a}(z)\},q_i\}\nonumber\\&&
- \int\textrm{d}y \ \delta v^{a_1}(y)\{q_i,\Phi_{a_1}(y)\} + \int\textrm{d}y \ \frac{d\epsilon^a(y)}{dt}*\{q_i,\Phi_a(y)\}=0.
\end{eqnarray}
Using the algebra (\ref{ve}) and (\ref{cc}), the above equation reduces to, 
\begin{eqnarray}
&&\int\textrm{d}z \ (\left[\frac{d\epsilon^b(z)}{dt}   - \int\textrm{d}y \ \epsilon^a(z)*[V_a^b(z,y)  + \int\textrm{d}u \ v^{a_1}(u)*C_{a_1a}^b(u,z,y)]\right]\nonumber\\&&*\frac{\partial\Phi_b(y)}{\partial p_i} - \delta v^{a_1}(z)*\frac{\partial\Phi_{a_1}(z)}{\partial p_i})
 = 0.\nonumber
\end{eqnarray}
Since the constraints are taken to be irreducible (i. e. independent) we get the following conditions, from the secondary and primary sectors, respectively,
\begin{eqnarray}
\frac{\textrm{d}\epsilon^{b_2}(x)}{\textrm{d}t}&=&\int \textrm{d}y \ \epsilon^a(y)*V^{b_2}_a(y,x)\nonumber\\
&&+\int \textrm{d}y \ \textrm{d}z \ \epsilon^a(y)*v^{a_1}(z)*C^{b_2}_{a_1a}(z,y,x)
\label{sl}
\end{eqnarray}
\begin{eqnarray}
\delta v^{b_1}(x)&=& \frac{d\epsilon^{b_1}(x)}{dt}   - \int\textrm{d}y \ \epsilon^a(y)*V_a^{b_1}(y,x) \nonumber\\
&&- \int\textrm{d}y \ \textrm{d}z \ \epsilon^a(y)*v^{a_1}(z)*C_{a_1a}^{b_1}(z,y,x).
\end{eqnarray}
The first relation expresses the fact that the gauge parameters $\epsilon^a$ are not all independent. In fact we find that, as stated earlier, the number of independent parameters of a gauge system is equal to the number of primary first class constraints. On the other hand, the second equation gives the variation of the Lagrange multipliers.

We will now use these results to analyse both star deformed gauge symmetries as well as twisted gauge symmetries.

\section{Star Deformed Gauge Symmetry}
So far we were discussing a general formulation for any gauge theory on noncommutative space. Now we concentrate on a particular model which describes a non-Abelian gauge field in the presence of a matter (fermionic) sector,
\begin{eqnarray}
S=\int \textrm{d}x \ [-\frac{1}{4}F_{\mu\nu}^a(x)*F^{\mu\nu a}(x)+\bar{\psi}(x)*(i\gamma^{\mu}D_{\mu}*-m)\psi(x)]
\label{L}
\end{eqnarray}
where
\begin{eqnarray}
&&D_{\mu}*\psi(x)\equiv \partial_{\mu}\psi(x)+igA_{\mu}(x)*\psi(x)\\
&&F_{\mu\nu}(x)\equiv\partial_{\mu}A_{\nu}(x)-\partial_{\nu}A_{\mu}(x)+ig[A_{\mu}(x),A_{\nu}(x)]_*.
\label{fmunu}
\end{eqnarray}
The above action is invariant under the star deformed gauge transformations,
\begin{eqnarray}
&&\delta_*A_{\mu}=\mathcal{D}_{\mu}*\eta=\partial_{\mu}\eta+ig(A_{\mu}*\eta-\eta*A_{\mu}),
\label{Amu}\\
&&\delta_*F_{\mu\nu}=ig[F_{\mu\nu},\eta]_*=ig(F_{\mu\nu}*\eta-\eta*F_{\mu\nu})
\label{Fmunu}\\
&&\delta_*\psi=-ig\eta*\psi
\label{si}\\
&&\delta_*\bar{\psi}=ig\bar{\psi}*\eta
\label{sibar}
\end{eqnarray}
with the usual Leibniz rule 
\begin{eqnarray}
\delta_*(A*B)=\delta_* A*B+A*\delta_* B.
\label{copro}
\end{eqnarray}
Varying the action (\ref{L}) with respect to the gauge field leads to the field equation
\begin{eqnarray}
\partial_{\mu}F^{\mu\nu}+ig[A_{\mu},F^{\mu\nu}]_*+j^{\nu}=0
\label{eqn}
\end{eqnarray}
where $j^{\nu}$ is the fermionic current
\begin{eqnarray}
j^{\nu}=g\psi_{\lambda}(\gamma^{\nu})_{\sigma\lambda}*\bar{\psi}_{\sigma}.
\end{eqnarray}
Note the ordering in which fermionic fields appear which is not equal to $-g\bar{\psi}_{\sigma}(\gamma^{\nu})_{\sigma\lambda}*\psi_{\lambda}$. This is due to the fact in calculating the variation of the term $\int \textrm{d}x \ \gamma^{\mu}\bar{\psi}*A_{\mu}*\psi$ we have used the cyclicity property of the star product,
\begin{eqnarray}
\int\textrm{d}x \ A*B*C=\int\textrm{d}x \ B*C*A=\int\textrm{d}x \ C*A*B
\label{cycl}
\end{eqnarray}
to keep $\delta_* A_{\mu}$ at an extreme end. We write the equation of motion (\ref{eqn}) in the form
\begin{eqnarray}
\mathcal{D}_{\mu}*F^{\mu\nu}+j^{\nu}=0
\end{eqnarray}
where
\begin{eqnarray}
&&\mathcal{D}_{\mu}*\xi=\partial_{\mu}\xi+ig[A_{\mu},\xi]_*
\end{eqnarray}
which, in component notation, reads, 
\begin{eqnarray}
&&(\mathcal{D}_{\mu}*\xi)^a=\partial_{\mu}\xi^a-\frac{g}{2}f^{abc}\{A_{\mu}^b,\xi^c\}_*+i\frac{g}{2}d^{abc}[A_{\mu}^b,\xi^c]_*
\label{D}
\end{eqnarray}
where the structure constants are defined by the symmetry matrices as,
\begin{eqnarray}
[T^a,T^b]=if^{abc}T^c
\label{sy1}\\
\{T^a,T^b\}=d^{abc}T^c.
\label{sy2}
\end{eqnarray}
The structure constants $f^{abc}$ and $d^{abc}$ can be made completely antisymmetric and completely symmetric as mentioned in \cite{rabin,amorim}.

We now start a Hamiltonian description of this theory. Throughout the paper we assume $\theta^{0i}=0$ to avoid higher order time derivatives. Due to the presence of grassmanian variables in our model (\ref{L}), the Poisson brackets in the previous section should be replaced by the graded brackets. The graded brackets between the fermionic variables are given by,
\begin{eqnarray}
\{\psi_{\alpha}(x),\psi_{\beta}^{\dagger}(y)\}=-i\delta_{\alpha\beta}\delta(x-y)
\label{psi}.
\end{eqnarray}
The canonical momentum of the Lagrangian (\ref{L}) is given by,
\begin{eqnarray}
\pi^a_{\sigma}=\frac{\partial\mathcal{L}}{\partial \dot{A^{\sigma a}}}=F^a_{\sigma 0}
\end{eqnarray}
which leads to a primary constraint 
\begin{eqnarray}
\Phi_1^a=\pi^a_0\approx0.
\label{con1}
\end{eqnarray}
The canonical Hamiltonian of the system is given by, 
\begin{eqnarray}
H&=&\int \textrm{d}x \  [\frac{1}{2}\pi^{ic}*\pi^{ic}+\frac{1}{4}F_{ij}^a*F^{ija}-(\mathcal{D}_i*\pi^i)^a*A_0^a\nonumber\\
&&-i\bar{\psi}*\gamma^i\partial_i\psi+g\bar{\psi}*\gamma^{\mu}A_{\mu}*\psi+m\bar{\psi}*\psi]
\label{ham}
\end{eqnarray}
where the operator $\mathcal{D}$ has already been defined in eq. (\ref{D}). Now using the basic Poisson bracket relation
\begin{eqnarray}
\{A^{\mu}(x),\pi_{\nu}(y)\}=\delta^{\mu}_{\nu}\delta(x-y)
\end{eqnarray}
the secondary constraints of the system are computed 
\begin{eqnarray}
&&\Phi_2^a=\{H,\Phi_1^a\}=\{H,\pi^a_0\}=(\mathcal{D}_i*\pi_i)^a-g\psi_{\lambda}* (T^a)_{\sigma\lambda}(\psi^{\dagger})_{\sigma}\approx 0
\label{con2}
\end{eqnarray}
where we have used
\begin{eqnarray}
\int \textrm{d}y \ A(y)*\delta(x-y)=\int \textrm{d}y \ A(y)\delta(x-y)=A(x).
\label{new1}
\end{eqnarray}
Note that this constraint is the zeroth component of the equation of motion (\ref{eqn}) expressed in phase space variables. The algebra of the $\Phi_1$ constraints is trivial,
\begin{eqnarray}
&&\{\Phi_1^a(x),\Phi_1^b(y)\}=0
\label{f11}\\
&&\{\Phi_1^a(x),\Phi_2^b(y)\}=0.
\label{f12}
\end{eqnarray}
The algebra of the constraint $\Phi_2$ with itself is also found to close, but in a nontrivial way. Since this calculation involves some subtleties, few intermediate steps are presented here. We write
\begin{eqnarray}
\Phi_2^a=T^a+\chi^a
\end{eqnarray}
where
\begin{eqnarray}
T^a&=&(\mathcal{D}_i*\pi_i)^a\nonumber\\
&=&\partial_{i}\pi^a_{i}-\frac{g}{2}f^{abc}\{A_{i}^b,\pi^{c}_{i}\}_*+i\frac{g}{2}d^{abc}[A_{i}^b,\pi^{c}_{i}]_* \  \ {\textrm{and}} \nonumber\\
\chi^a&=&-g\psi_{\lambda}* (T^a)_{\sigma\lambda}(\psi^{\dagger})_{\sigma}
\end{eqnarray}
The graded brackets of the terms $T^a$ and $\chi^a$ separately close among themselves. Let us show it first for $T^a$\cite{amorim}. Using the identity\cite{rabin,amorim}
\begin{eqnarray}
A(x)*\delta(x-y)=\delta(x-y)*A(y)
\label{id}
\end{eqnarray}
we obtain
\begin{eqnarray}
\{\partial_i\pi^{a}_i(x),-\frac{g}{2}f^{bcd}\{A_j^c(y),\pi^{d}_{j}(y)\}_*\}+\{-\frac{g}{2}f^{acd}\{A_i^c(x),\pi^{d}_{i}(x)\}_*,\partial_j\pi^{b}_j(y)\}\nonumber\\=\frac{g}{2}f^{abc}\{\delta(x-y),\partial_i\pi^{c}_{i}(x)\}_*
\label{1z}
\end{eqnarray}
and
\begin{eqnarray}
\{\partial_i\pi^{a}_i(x),i\frac{g}{2}d^{bcd}[A_j^c(y),\pi^{d}_{j}(y)]_*\}+\{i\frac{g}{2}d^{acd}[A_i^c(x),\pi^{d}_{i}(x)]_*,\partial_j\pi_j^{b}(y)\}\nonumber\\
=-i\frac{g}{2}d^{abc}[\delta(x-y),\partial_i\pi^{c}_i(x)]_*
\label{2z}
\end{eqnarray}
Exploiting the Jacobi identity 
\begin{eqnarray}
[\pi_i(x),[A_i(x),\, T^b\delta(x-y)]_*]_*+[A_i(x),[T^b\delta(x-y) ,\pi_i(x)]_*]_*\nonumber\\
+[T^b\delta(x-y),[\pi_i(x),A_i(x)]_*]_*=0
\end{eqnarray}
the remaining terms of $\{T^a(x),T^b(y)\}$ are written as
\begin{equation}
\frac{i}{2}g^2f^{abc}\{\delta(x-y),[A_i,\pi_i]_*^c\}_*+\frac{1}{2}g^2d^{abc}[\delta(x-y),[A_i,\pi_i]_*^c]_*.
\label{3z}
\end{equation}

Combining the expressions (\ref{1z}), (\ref{2z}) and (\ref{3z}), we get the closed algebra
\begin{equation}
\{T^a(x),T^b(y)\}=\frac{g}{2}f^{abc}\{\delta(x-y),T^c(x)\}_*-
i\frac{g}{2}d^{abc}[\delta(x-y),T^c(x)]_*
\label{T}
\end{equation}
Now to show that the graded bracket $\{\chi^a(x),\chi^b(y)\}$ really closes we use the product rule
\begin{eqnarray}
\begin{array}{rcl}
\{A,BC\}=\{A,B\}C+(-1)^{\eta_A\eta_B}B\{A,C\}\\
\{AB,C\}=A\{B,C\}+(-1)^{\eta_B\eta_C}\{A,C\}B
\end{array}
\label{pro}
\end{eqnarray}
where
\begin{eqnarray}
&&\eta=0 \  \  \ {\textrm{for bosonic variable and}}\nonumber\\
&&\eta=1 \  \  \ {\textrm{for fermionic variable}}\nonumber
\end{eqnarray}
Eq. (\ref{pro}), together with the bracket (\ref{psi}) and the identity (\ref{id}), yields,
\begin{eqnarray}
\{\chi^a(x),\chi^b(y)\}=\frac{g}{2}f^{abc}\{\delta(x-y),\chi^c(x)\}_*-i\frac{g}{2}d^{abc}[\delta(x-y),\chi^c(x)]_*.
\label{ch}
\end{eqnarray}
Thus eqs. (\ref{T}) and (\ref{ch}) imply the closure of $\Phi_2$,
\begin{eqnarray}
\{\Phi^a_2(x),\Phi^b_2(y)\}=\frac{g}{2}f^{abc}\{\delta(x-y),\Phi^c_2(x)\}_*-i\frac{g}{2}d^{abc}[\delta(x-y),\Phi^c_2(x)]_*.
\end{eqnarray}
Likewise the involutive algebra of the canonical Hamiltonian with the constraints is found to be,
\begin{eqnarray}
&&\{H_c,\Phi_1^a\}=\Phi_2^a
\label{al}\\
&&\{H_c,\Phi_2^a\}=-\frac{g}{2}f^{abc}\{A^{0b},\Phi_2^c\}_*+i\frac{g}{2}d^{abc}[A^{0b},\Phi_2^c]_*.
\label{al1}
\end{eqnarray}
Due to the algebra (\ref{f11}) and (\ref{f12}) the term $C^{b_2}_{a_1a}$ in the r. h. s. of eq. (\ref{cc}) vanishes. So we simplify eq. (\ref{sl}) as
\begin{eqnarray}
\frac{\textrm{d}\epsilon^{b_2}(x)}{\textrm{d}t}=\int \textrm{d}y \ \epsilon^a(y)*V^{b_2}_a(y,x).
\label{sl1}
\end{eqnarray}
The $V$ function defined in eq. (\ref{ve}) can be found from the algebra (\ref{al}) and (\ref{al1}) as
\begin{eqnarray}
(V^2_1)^{ab}(x,y)&=&\delta^{ab}\delta(x-y),
\label{50}\\
(V^2_2)^{ab}(x,y)&=&\frac{g}{2}f^{abc}\{\delta(x-y),A^{0c}(y)\}_*\nonumber\\
&&+i\frac{g}{2}d^{abc}[\delta(x-y),A^{0c}(y)]_*.
\label{51}
\end{eqnarray}
Now we write eq. (\ref{sl1}) in its expanded form as,
\begin{eqnarray}
\frac{\textrm{d}\epsilon^{2a}(x)}{\textrm{d}t}=\int \textrm{d}y \ \epsilon^{1b}(y)*(V^2_1)^{ba}(y,x)+\int \textrm{d}y \ \epsilon^{2b}(y)*(V^2_2)^{ba}(y,x).
\end{eqnarray}
Using (\ref{50}) and (\ref{51}) in the above eq. we get
\begin{eqnarray}
\dot \epsilon^{2a}=\epsilon^{1a}-\frac{g}{2}f^{abc}\{\epsilon^{2b}(x),A^{0c}(x)\}_*+i\frac{g}{2}d^{abc}[\epsilon^{2b}(x),A^{0c}(x)]_*
\end{eqnarray}
so that
\begin{eqnarray}
\epsilon^{1a}=(\mathcal{D}_0*\epsilon^2)^a.
\label{54}
\end{eqnarray}
Thus, using the above result, the generator given in (\ref{G6}) is written in terms of a single parameter as,
\begin{eqnarray}
G=\int \textrm{d}x \ (\mathcal{D}_0*\epsilon^{2})^a*\Phi_1^a+\epsilon^{2a}*\Phi_2^a
\label{generator}
\end{eqnarray}
where the constraints $\Phi_1$ and $\Phi_2$ are defined in (\ref{con1}) and (\ref{con2}). After obtaining the complete form of the generator, we are now in a position to calculate the variation of the different fields. The general formula to get this follows from (\ref{gt}),
\begin{eqnarray}
\delta q_{\alpha}(x)&=&\int \textrm{d}y \ \epsilon^b(y)*\{q_{\alpha}(x), \Phi_b(y)\}, \ b=1,2.
\label{eps1}\\
&=&\int \textrm{d}y \ (\mathcal{D}_0*\epsilon^{2})^a(y)*\{q_{\alpha}(x), \Phi^a_1(y)\}\nonumber\\
&&+\int \textrm{d}y \ \epsilon^{2a}(y)*\{q_{\alpha}(x), \Phi^a_2(y)\}.
\label{eps2}
\end{eqnarray}
Let us first study the gauge transformation of the field $A^{\mu}$. The variation of its time component is
\begin{eqnarray}
\delta_* A^a_{0}(x)&=&\int \textrm{d}y \ (\mathcal{D}_0*\epsilon^{2})^b(y)*\{A^a_{0}(x),\pi^b_0(y)\}\nonumber\\
&=&\int \textrm{d}y \ (\mathcal{D}_0*\epsilon^{2})^b(y)\delta^{ab}*\delta(x-y)\nonumber\\
&=&\int \textrm{d}y \ (\mathcal{D}_0*\epsilon^{2})^a(y)\delta(x-y)\nonumber\\
&=&(\mathcal{D}_0*\epsilon^{2})^{a}
\label{A0}
\end{eqnarray}
where we have used the identity (\ref{new1}). The variation of the space component is likewise given by,
\begin{eqnarray}
\delta_* A^a_{i}(x)&=&\int \textrm{d}y \ \epsilon^{2b}(y)*\{A^a_{i}(x),\mathcal{D}_j*\pi_j^b(y)\}\nonumber\\
&=&\int \textrm{d}y \ \epsilon^{2b}(y)*(-\partial_i^y\delta(x-y)\delta^{ab}+\frac{g}{2}f^{bca}\{A_i^c(y),\delta(x-y)\}_*\nonumber\\
&&-i\frac{g}{2}d^{bca}[A_i^c(y),\delta(x-y)]_*).\nonumber
\end{eqnarray}
Now dropping the boundary term and using the cyclicity property (\ref{cycl}) we write the above expression as
\begin{eqnarray}
\delta_* A^a_{i}(x)&=&\partial_i\epsilon^{2a}-\frac{g}{2}f^{abc}\{A_i^b,\epsilon^{2c}\}_*+i\frac{g}{2}d^{abc}[A_i^b,\epsilon^{2c}]_*\nonumber\\
&=&(\mathcal{D}_i*\epsilon^{2})^a(x).
\label{49}
\end{eqnarray}
Combining eqs. (\ref{A0}) and (\ref{49}) we obtain,
\begin{eqnarray}
\delta_* A_{\mu}^a=(\mathcal{D}_{\mu}*\epsilon^2)^a
\end{eqnarray}
thereby reproducing (\ref{Amu}) with the identification $\epsilon^2\rightarrow\eta$. In a likewise manner the gauge transformation of the matter fields is also obtained,
\begin{eqnarray}
\delta_*\psi_{\alpha}(x)&=&\int\textrm{d}y \ \epsilon^{2a}(y)*\{\psi_{\alpha}(x),\Phi^{2a}(y)\}\nonumber\\
&=&\int\textrm{d}y \ \epsilon^{2a}(y)*\{\psi_{\alpha}(x),-g\psi_{\lambda}(y)*(T^a)_{\sigma\lambda}\psi_{\sigma}^{\dagger}(y)\nonumber\\
&=&\int\textrm{d}y \ \epsilon^{2a}(y)*\psi_{\lambda}(y)*(T^a)_{\sigma\lambda}(-i)\delta_{\alpha \sigma}\delta(x-y)
\label{fr}
\end{eqnarray}
where (\ref{psi}) has been used. Now using the property (\ref{new1}), the above equation is written as
\begin{eqnarray}
\delta_*\psi_{\alpha}(x)=-ig\epsilon^{2a}(x)*\left(T^a\right)_{\alpha\beta}\psi_{\beta}(x).
\label{deltsi}
\end{eqnarray}
In a similar way we get
\begin{eqnarray}
\delta_*\bar{\psi}_{\alpha}(x)&=&\int\textrm{d}y \ \epsilon^{2a}(y)*\{\bar{\psi}_{\alpha}(x),\Phi^{2a}(y)\}\nonumber\\
&&=ig\left(T^a\right)_{\beta\alpha}\bar{\psi}_{\beta}(x)*\epsilon^{2a}(x)
\label{deltsibar}
\end{eqnarray}
which reproduces (\ref{si}) and (\ref{sibar}).

It is also possible to compute the gauge variations of star composites in the same way. For example,
\begin{eqnarray}
\delta_*(\psi_{\alpha}(x)*\psi_{\beta}(x))&=&\int\textrm{d}y \ \epsilon^{2a}(y)*\{\psi_{\alpha}(x)*\psi_{\beta}(x),\Phi^{2a}(y)\}\nonumber\\ 
&=&ig\int\textrm{d}y \ (T^a)_{\beta \lambda}\epsilon^{2a}(y)*\psi_{\lambda}(y)*\psi_{\alpha}(x)*\delta(x-y)\nonumber\\&&-ig\int\textrm{d}y \ (T^a)_{\alpha \lambda}\epsilon^{2a}(y)*\psi_{\lambda}(y)*\delta(x-y)*\psi_{\beta}(x).\nonumber
\end{eqnarray}
Using the identity (\ref{id}) the argument of $\psi_{\alpha}$ in the first integral and that of $\psi_{\beta}$ in the second integral is shifted from $x$ to $y$ so that star product is defined only at the same point ($y$). Finally, using (\ref{cycl}) and (\ref{new1}), and keeping in mind the grassmanian nature of the fermionic field we get
\begin{eqnarray}
\delta_*(\psi_{\alpha}*\psi_{\beta})=-ig\left((T^a)_{\beta \lambda}\psi_{\alpha}*\epsilon^{2a}*\psi_{\lambda}+(T^a)_{\alpha \lambda}\epsilon^{2a}*\psi_{\lambda}*\psi_{\beta}\right).
\end{eqnarray}
This is the result one also finds by using (\ref{deltsi}) and the standard coproduct rule,
\begin{eqnarray}
\delta_*(\psi_{\alpha}*\psi_{\beta})&=&(\delta_*\psi_{\alpha})*\psi_{\beta}+\psi_{\alpha}*(\delta_*\psi_{\beta})\\
&=&-ig\left(\psi_{\alpha}*\epsilon^{2a}(T^a)_{\beta \lambda}*\psi_{\lambda}+\epsilon^{2a}(T^a)_{\alpha \lambda}*\psi_{\lambda}*\psi_{\beta}\right)
\end{eqnarray}
other star composites can be treated identically. This culminates our analysis of star deformed gauge symmetry. Note that the standard coproduct rule (\ref{copro}) is necessary for the invariance of the action as well as the consistency of the analysis.
\section{Twisted Gauge Symmetry}
So far we were discussing about the star deformed gauge transformation from a general Hamiltonian formulation which obeys the normal coproduct rule (\ref{copro}). But as discussed in \cite{wess,vas,chai,jwess} the action (\ref{L}) is also invariant under the undeformed gauge transformations
\begin{eqnarray}
\begin{array}{rcl}
&&\delta_{\eta} A_{\mu}=\mathcal{D}_{\mu}\eta=\partial_{\mu}\eta+ig(A_{\mu}\eta-\eta A_{\mu}),\\
&&\delta_{\eta} F_{\mu\nu}=ig[F_{\mu\nu},\eta]=ig(F_{\mu\nu}\eta-\eta F_{\mu\nu})\\
&&\delta_{\eta} \psi=-ig\eta\psi\\
&&\delta_{\eta} \bar{\psi}=ig\bar{\psi}\eta
\end{array}
\label{YY}
\end{eqnarray}
with the twisted Leibniz rule\cite{wess,vas,jwess},
\begin{eqnarray}
\delta_{\eta}(f*g)&=&\sum_n(\frac{-i}{2})^n\frac{\theta^{\mu_1\nu_1}\cdot \cdot \cdot\theta^{\mu_n\nu_n}}{n!}\nonumber\\
&&(\delta_{\partial_{\mu_1}\cdot \cdot \cdot\partial_{\mu_n}\eta}f*\partial_{\nu_1}\cdot \cdot \cdot\partial_{\nu_n}g+\partial_{\mu_1}\cdot \cdot \cdot\partial_{\mu_n}f*\delta_{\partial_{\nu_1}\cdot \cdot \cdot\partial_{\nu_n}\eta}g).
\label{co}
\end{eqnarray}

This rule is also essential to obtain $\delta_{\eta}F_{\mu\nu}$. Using (\ref{fmunu}) and,
\begin{eqnarray}
\delta_{\eta}(A_{\mu}*A_{\nu})&=&\partial_{\mu}\eta A_{\nu}+A_{\mu}\partial_{\nu}\eta-ig\eta^a\left([T^a,A_{\mu}]*A_{\nu}+A_{\mu}*[T^a,A_{\nu}]\right)
\nonumber\\
&=&\partial_{\mu}\eta A_{\nu}+A_{\mu}\partial_{\nu}\eta-ig[\eta,(A_{\mu}*A_{\nu})]
\label{new2}
\end{eqnarray}
following from (\ref{co}), immediately leads to the undeformed transformation (\ref{YY}) for $\delta_{\eta}F_{\mu\nu}$. The gauge variation of the other star composites are similarly computed from (\ref{co}),
\begin{eqnarray}
&&\delta_{\eta}(A_{\mu}*\psi)=(\partial_{\mu}\eta)\psi-ig\eta(A_{\mu}*\psi)
\label{g.v.}\\
&&\delta_{\eta}(\phi*\psi)=-ig\eta^a\left((T^a\phi)*\psi+\phi*(T^a\psi)\right).
\label{73}
\end{eqnarray}
In an analogous manner we can also obtain the gauge variation of a chain of fields, as for example
\begin{eqnarray}
\delta_{\eta}(\phi*\psi*\chi)&=&-ig\eta^a((T^a\phi)*\psi*\chi+\phi*(T^a\psi)*\chi\nonumber\\&&+\phi*\psi*(T^a\chi))
\label{3f}\\
\delta_{\eta}(\phi*\psi*A_{\mu})&=&-ig\eta^a\left((T^a\phi)*\psi*A_{\mu}+\phi*(T^a\psi)*A_{\mu}\right)\nonumber\\
&&+ig\eta^a\phi*\psi*[A_{\mu},T^a]+(\phi*\psi)\partial_{\mu}\eta
\label{2f}
\end{eqnarray}
where $\phi$ and $\chi$ have similar transformation properties as $\psi$. We observe that the transformation rules for the star products of variables is also identical to the corresponding undeformed relations, as for example,
\begin{eqnarray}
\delta_{\eta}(A_{\mu}A_{\nu})=\partial_{\mu}\eta A_{\nu}+A_{\mu}\partial_{\nu}\eta-ig[\eta,(A_{\mu}A_{\nu})]
\end{eqnarray}
where $A_{\mu}$ is the commutative space gauge field with normal gauge transformation.

We now present an alternative interpretation of the twisted coproduct rule (\ref{co}). The results (\ref{new2}), (\ref{g.v.}), (\ref{73}) and also (\ref{3f}), (\ref{2f}) are seen to follow by using the standard coproduct rule (\ref{copro}) but pushing the gauge parameter $\eta$ outside the star operation at the end of the computations. Denoting this manipulation as,
\begin{eqnarray}
\delta_{\eta}(A*B)\sim(\delta_{\eta}A)*B+A*(\delta_{\eta}B)
\end{eqnarray}
we find
\begin{eqnarray}
\delta_{\eta}(\phi*\psi)&\sim&(\delta_{\eta}\phi)*\psi+\phi*(\delta_{\eta}\psi)\\
&\sim&-ig(\eta\phi)*\psi-ig\phi*(\eta\psi)\\
&=&-ig\eta^a\{(T^a\phi)*\psi+\phi*(T^a\psi)\}
\end{eqnarray}
which reproduces (\ref{73}). Likewise we see,
\begin{eqnarray}
\delta_{\eta}(A_{\mu}*\psi)&\sim&(\delta_{\eta}A_{\mu})*\psi+A_{\mu}*(\delta_{\eta}\psi)\nonumber\\
&\sim&(\partial_{\mu}\eta-ig\eta^a[T^a,A_{\mu}])*\psi+A_{\mu}*(-ig\eta^aT^a\psi)\nonumber\\
&=&\partial_{\mu}\eta\psi-ig\eta^a([T^a,A_{\mu}]*\psi)-ig\eta^a(A_{\mu}*T^a\psi)\nonumber\\
&=&\partial_{\mu}\eta\psi-ig\eta(A_{\mu}*\psi)
\end{eqnarray}
which reproduces (\ref{g.v.}). Similarly, 
\begin{eqnarray}
\delta_{\eta}(\phi*\psi*\chi)&\sim&(\delta_{\eta}\phi)*\psi*\chi+\phi*(\delta_{\eta}\psi)*\chi+\phi*\psi*(\delta_{\eta}\chi)\\
&\sim&-ig(\eta\phi)*\psi*\chi-ig\phi*(\eta\psi)*\chi-ig\phi*\psi*(\eta\chi)\\
&=&-ig\eta^a\{(T^a\phi)*\psi*\chi+\phi*(T^a\psi)*\chi\nonumber\\
&&+\phi*\psi*(T^a\chi)\}
\end{eqnarray}
thereby reproducing (\ref{3f}).

We now suitably modify the Hamiltonian formulation of the previous section to systematically obtain the undeformed gauge transformations (\ref{YY}) as well as the relations (\ref{new2}), (\ref{g.v.}), (\ref{73}) manifesting the twisted Leibniz rule. As far as the gauge generator is concerned the analysis is similar to the previous case and the same expression (\ref{generator}) is obtained. This is not unexpected since the Gauss constraint defining the generator is basically the time component of the field equations which are identical in both treatments. The difference can come only through the computation of the relevant Poisson brackets that lead to the gauge transformations. In our interpretation the twisted coproduct is just the standard coproduct with the proviso that the gauge parameter is pushed outside the star operation at the end of the computations. We adopt a similar prescription for computing the modified Poisson brackets.

 The gauge variation of the time component of $A^{\mu}$ field is found by suitably Poisson bracketing with (\ref{generator}) (renaming $\epsilon^{2}$ as $\eta$),
\begin{eqnarray}
\delta_{\eta} A^a_{0}(x)&=&\int \textrm{d}y \ (\mathcal{D}_0*\eta)^b(y)*\{A^a_{0}(x),\pi^b_0(y)\}\nonumber\\
&\sim&\int \textrm{d}y \ (\mathcal{D}_0*\eta)^b(y)\delta^{ab}*\delta(x-y)\nonumber\\
&\sim&\int(\textrm{d}y \ \partial_0\eta^{a}-\frac{g}{2}f^{abc}\{A_0^b,\eta^{c}\}_*+i\frac{g}{2}d^{abc}[A_0^b,\eta^{c}]_*)(y)*\delta(x-y)\nonumber\\
&=&\partial_0\eta^{a}-gf^{abc}A_0^b\eta^{c}\nonumber
\end{eqnarray}
where in the last step we put $\eta$ outside the star product following our prescription. This is written in a compact notation as,
\begin{eqnarray}
\delta_{\eta} A^a_{0}=(\mathcal{D}_0\eta)^a.
\label{tw1}
\end{eqnarray}
The variation of the space component is also calculated in a similar way
\begin{eqnarray}
\delta_{\eta} A^a_{i}(x)&=&\int \textrm{d}y \ \eta^{b}(y)*\{A^a_{i}(x),\mathcal{D}_j*\pi_j^b(y)\}\nonumber\\
&\sim&\int \textrm{d}y \ \eta^{b}(y)*(-\partial_i^y\delta(x-y)\delta^{ab}+\frac{g}{2}f^{bca}\{A_i^c(y),\delta(x-y)\}_*\nonumber\\
&&-i\frac{g}{2}d^{bca}[A_i^c(y),\delta(x-y)]_*)\nonumber\\
&\sim&\int \textrm{d}y \ \eta^{a}(y)*(-\partial_i^y\delta(x-y))+\nonumber\\&&\frac{g}{2}f^{bca}(\eta^{b}(y)*A_i^c(y)*\delta(x-y)+\eta^{b}(y)*\delta(x-y)*A_i^c(y))\nonumber\\
&&-i\frac{g}{2}d^{bca}(\eta^{b}(y)*A_i^c(y)*\delta(x-y)-\eta^{b}(y)*\delta(x-y)*A_i^c(y))\nonumber
\end{eqnarray}
Now dropping the boundary term, using the cyclicity property (\ref{cycl}) and the relation (\ref{new1}) we write the above expression as
\begin{eqnarray}
\delta_{\eta} A^a_{i}(x)&\sim&\partial_i\eta^{a}(x)+\nonumber\\&&\frac{g}{2}f^{bca}(\eta^{b}(x)*A_i^c(x)+A_i^c(x)*\eta^{b}(x))\nonumber\\
&&-i\frac{g}{2}d^{bca}(\eta^{b}(x)*A_i^c(x)-A_i^c(x)*\eta^{b}(x))\nonumber
\end{eqnarray}
Finally, keeping the gauge parameter $\eta$ outside the star product we obtain
\begin{eqnarray}
\delta_{\eta} A^a_{i}(x)&=&\partial_i\eta^{a}-gf^{abc}A_i^b\eta^{c}\nonumber\\
&=&(\mathcal{D}_i\eta)^{a}(x).
\label{tw2}
\end{eqnarray}
Combining eqs. (\ref{tw1}) and (\ref{tw2}) we write the gauge variation in a covariant notation
\begin{eqnarray}
\delta_{\eta} A_{\mu}^a=(\mathcal{D}_{\mu}\eta)^a
\label{77}
\end{eqnarray}

The gauge variation of the fermionic field can be obtained in a similar way
\begin{eqnarray}
&&\delta_{\eta}\psi_{\alpha}(x)=-ig\eta^{a}(x)\left(T^a\right)_{\alpha\beta}\psi_{\beta}(x)
\label{78}\\
&&\delta_{\eta}\bar{\psi}_{\alpha}(x)=ig\left(T^a\right)_{\beta\alpha}\bar{\psi}_{\beta}(x)\eta^{a}(x).
\label{79}
\end{eqnarray}
The calculation of the gauge variation of composite fields needs some care. For example we consider the variation $\delta_{\eta}(A_{\mu}*\psi)$,
\begin{eqnarray}
\delta_{\eta}(A_{0}(x)*\psi(x))&=&T^a\delta(A_{0}^a(x)*\psi(x))\nonumber\\
&\sim&T^a\int \textrm{d}y \ (\mathcal{D}_0*\eta^{b})(y)*\{A^a_{0}(x)*\psi(x),\pi^b_0(y)\}+\nonumber\\
&&T^b\int\textrm{d}y \ \eta^{c}(y)*\{A^b_{0}(x)*\psi(x),-g\psi(y)*(T^c)\psi^{\dagger}(y)\}\nonumber\\
&\sim&T^a\int \textrm{d}y \ (\mathcal{D}_0*\eta^{a})(y)*\delta(x-y)*\psi(x)\nonumber\\
&&-igT^b\int\textrm{d}y \ \eta^{c}(y)*T^c\psi(y)*A^b_{0}(x)*\delta(x-y).
\end{eqnarray}
As mentioned earlier, the star product is defined only at the same point of two functions. So to evaluate the above integral we use the identity (\ref{id}) to change the argument of $\psi$ and $A^b_{0}$ from $x$ to $y$. Thus we obtain
\begin{eqnarray}
\delta_{\eta}(A_{0}(x)*\psi(x))&\sim&T^a\int \textrm{d}y \ (\mathcal{D}_0*\eta^{a})(y)*\psi(y)*\delta(x-y)\nonumber\\
&&-igT^b\int\textrm{d}y \ \eta^{c}(y)T^c*\psi(y)*\delta(x-y)*A^b_{0}(y).
\end{eqnarray}
Using the properties (\ref{cycl}), (\ref{new1}) and finally removing the gauge parameter $\eta$ outside the star product we obtain
\begin{eqnarray}
\delta_{\eta}(A_{0}*\psi)&=&T^a(\partial_0\eta^{a}\psi-gf^{abc}\eta^{c}(A_{0}^b*\psi))-igT^bT^c\eta^{c}(A_{0}^b*\psi).
\end{eqnarray}
Following the symmetry properties (\ref{sy1},\ref{sy2}) we write the above result as
\begin{eqnarray}
\delta_{\eta}(A_{0}*\psi)&=&T^a(\partial_0\eta^{a}\psi-gf^{abc}\eta^{c}(A_{0}^b*\psi))\nonumber\\
&&-igT^a\eta^{c}(A_{0}^b*\psi)(\frac{1}{2}d^{bca}+\frac{i}{2}f^{bca})\\
&=&T^a(\partial_0\eta^{a}\psi)+gT^a\eta^{c}(A_{0}^b*\psi)(-\frac{i}{2}d^{bca}+\frac{1}{2}f^{bca}).
\label{ta0}
\end{eqnarray}
The space part is also obtained in a similar way
\begin{eqnarray}
\delta_{\eta}(A_{i}*\psi)=T^a(\partial_i\eta^{a}\psi)+gT^a\eta^{c}(A_{i}^b*\psi)(-\frac{i}{2}d^{bca}+\frac{1}{2}f^{bca}).
\label{tai}
\end{eqnarray}
Expressions (\ref{ta0}, \ref{tai}) are basically the time and space component of the equation (\ref{g.v.}). Finally, we calculate the gauge variation of a star product of three fields,
\begin{eqnarray}
\delta_{\eta}(\psi_{\alpha}*\psi_{\beta}*\psi_{\gamma})(x)&\sim&\int\textrm{d}y \ \eta(y)*\{\psi_{\alpha}(x)*\psi_{\beta}(x)*\psi_{\gamma}(x), -g\psi_{\lambda}(y)*T^a_{\sigma\lambda}\psi_{\sigma}^{\dagger}\}\nonumber\\
&\sim&-ig\int\textrm{d}y \ \eta(y)*T^a_{\gamma \lambda}\psi_{\lambda}(y)*\psi_{\alpha}(x)*\psi_{\beta}(x)*\delta(x-y)\nonumber\\
&&+ig\int\textrm{d}y \ \eta(y)*T^a_{\beta \lambda}\psi_{\lambda}(y)*\psi_{\alpha}(x)*\delta(x-y)*\psi_{\gamma}(x)\nonumber\\
&&-ig\int\textrm{d}y \ \eta(y)*T^a_{\alpha \lambda}\psi_{\lambda}(y)*\delta(x-y)*\psi_{\beta}(x)*\psi_{\gamma}(x)\nonumber
\end{eqnarray}
Now using the property (\ref{id}) and its generalisation,
\begin{eqnarray}
A(x)*\delta(x-y)*B(x)=B(y)*\delta(x-y)*A(y)
\end{eqnarray}
all arguments of the above equation are shifted to $y$ to yield,
\begin{eqnarray}
\delta_{\eta}(\psi_{\alpha}*\psi_{\beta}*\psi_{\gamma})(x)&\sim&-ig\int\textrm{d}y \ \eta(y)*T^a_{\gamma \lambda}\psi_{\lambda}(y)*\delta(x-y)*\psi_{\alpha}(y)*\psi_{\beta}(y)\nonumber\\
&&-ig\int\textrm{d}y \ \eta(y)*T^a_{\beta \lambda}\psi_{\lambda}(y)*\psi_{\gamma}(y)*\delta(x-y)*\psi_{\alpha}(y)\nonumber\\
&&-ig\int\textrm{d}y \ \eta(y)*T^a_{\alpha \lambda}\psi_{\lambda}(y)*\psi_{\beta}(y)*\psi_{\gamma}(y)*\delta(x-y)\nonumber
\end{eqnarray}
where the extra negative sign in the second integral is due to the flip of two grassmanian variables. Making use of the identities (\ref{cycl}) and (\ref{new1}) we evaluate the above integral and finally removing the gauge parameter $\eta$ outside the star product we obtain
\begin{eqnarray}
\delta_{\eta}(\psi_{\alpha}*\psi_{\beta}*\psi_{\gamma})&=&-ig\eta^a\{(T^a\psi)_{\alpha}*\psi_{\beta}*\psi_{\gamma}+\psi_{\alpha}*(T^a\psi)_{\beta}*\psi_{\gamma}\nonumber\\
&&+\psi_{\alpha}*\psi_{\beta}*(T^a\psi)_{\gamma}\}
\end{eqnarray}
which is just eq. (\ref{3f}) written in component form. The gauge variations of the other composites are computed in the same way reproducing the results (\ref{new2}), (\ref{73}), (\ref{2f}) obtained by using the twisted coproduct rule.

This section is concluded by making a comparison with results obtained in \cite{jwess} using Hopf algebra techniques\cite{abe}. In this approach the gauge generator (in the Schroedinger representation) is taken exactly as in the undeformed situation,
\begin{eqnarray}
G_{\alpha}=\int d\textrm{z} \ \left(\partial_{\mu}\alpha^l(z)+g\alpha^r(z)A_{\mu}^s(z)f^{rsl}\right)\frac{\delta}{\delta A_{\mu}^l(z)}
\label{w1}
\end{eqnarray}
which is consistent with the algebra,
\begin{eqnarray}
[G_{\alpha},G_{\beta}]=igG_{[\alpha,\beta]}.
\label{w2}
\end{eqnarray}
The usual coproduct is then twisted by introducing the generator (\ref{w1}) and is shown to be compatible with the general expression (\ref{co}).

It should however be pointed out that the generator (\ref{w1}) only generates the undeformed gauge transformations. Star deformed gauge transformations are obviously not generated by it.

In our unified approach the generator is given by (\ref{generator}). Depending on the interpretation of computing the Poisson brackets of this generator with the field variables yields either the star deformed gauge transformations or the undeformed gauge transformations with the twisted coproduct. The generator (\ref{generator}) satisfies a star deformed version of (\ref{w2})
\begin{eqnarray}
[G_{\alpha},G_{\beta}]=gG_{[\alpha,\beta]_*}.
\end{eqnarray}
\section{Conclusions}
The conclusion of our work is that, as far as gauge symmetry is concerned, both star deformed symmetry and twisted symmetry are on an equal footing. The gauge generator, obtained in the Hamiltonian formalism, reproduced star deformed gauge transformations with a normal coproduct as well as undeformed gauge transformations with a twisted coproduct. This was based on an appropriate interpretation of computing the Poisson brackets that led to the gauge transformations.

The present analysis revealed a new interpretation of the twisted coproduct. It was found that the twisted coproduct was equivalent to the normal coproduct with the condition that the gauge parameter had to be taken outside the star operation at the end of the computations.

A point which has been stressed in the literature\cite{jwess,gaume} is that twisted symmetry is not a physical symmetry in the usual sense and it is uncertain whether Noether charges and ward identities can be obtained. This is because twisted invariance leads to transformations that do not act only on the fields. Nevertheless we were successful in suitably defining gauge generators and transformations. This was quite reassuring since for a genuine symmetry (twisted or otherwise), a generator must be appropriately defined. We feel this to be an important step in regarding a deformed gauge theory as a theory with properties similar to what we desire for physics.

\end{document}